\documentclass[twocolumn]{aastex62}

\newcommand{\msun}{{\rm M}_\odot}

\newcommand{\lsun}{{\rm L}_\odot}
\newcommand{\cc}{{\rm cm}^{-3}}

\newcommand{\cmpc}{{\rm cMpc}}

\newcommand{\pc}{{\rm pc}}
\newcommand{\kms}{{\rm km~s}^{-1}}
\newcommand{\hz}{{\rm Hz}}

\newcommand{\beq}{\begin{equation}}
\newcommand{\eeq}{\end{equation}}

\received{\today}
\revised{\today}
\accepted{\today}
\submitjournal{ApJ}

\shorttitle{gw from merging smbh in ulirg}
\shortauthors{Inayoshi, Ichikawa \& Haiman}

\begin{document}

\title{Gravitational waves from supermassive black hole binaries in ultra-luminous infrared galaxies}

\email{inayoshi@astro.columbia.edu (KI)}

\author{Kohei Inayoshi}
\affil{Department of Astronomy, Columbia University, 550 W. 120th Street, New York, NY 10027, USA}

\author{Kohei Ichikawa}
\affiliation{Department of Astronomy, Columbia University, 550 W. 120th Street, New York, NY 10027, USA}
\affiliation{Frontier Research Institute for Interdisciplinary Sciences, Tohoku University, Sendai 980-8578, Japan}

\author{Zolt\'an Haiman}
\affiliation{Department of Astronomy, Columbia University, 550 W. 120th Street, New York, NY 10027, USA}

\begin{abstract}
Gravitational waves (GWs) in the nano-hertz band are great tools for understanding
the cosmological evolution of supermassive black holes (SMBHs) in galactic nuclei.
We consider SMBH binaries in high-$z$
ultra-luminous infrared galaxies (ULIRGs) as sources of a stochastic
GW background (GWB). ULIRGs are likely associated with gas-rich galaxy
mergers containing SMBHs that possibly occur at most once in the life
of galaxies, unlike multiple dry mergers at low redshift.
Adopting a well-established sample of ULIRGs,
we study the properties of the GWB due to coalescing binary SMBHs in these galaxies. 
Since the ULIRG population peaks at $z>1.5$, the amplitude of the GWB is not affected 
even if BH mergers are delayed by as long as $\sim $ 10 Gyrs.
Despite the rarity of the high-$z$ ULIRGs, 
we find a tension with the upper limits from Pulsar Timing Array (PTA) experiments.
This result suggests that if a fraction $f_{\rm m,gal}$ of ULIRGs are associated with SMBH binaries, then
no more than $20 f_{\rm m,gal}(\lambda_{\rm Edd}/0.3)^{5/3}(t_{\rm life}/30~{\rm Myr})~\%$
of the binary SMBHs in ULIRGs can merge within a Hubble time,  
for plausible values of the Eddington ratio of ULIRGs ($\lambda_{\rm Edd}$) and their lifetime ($t_{\rm life}$).
\end{abstract}


\section{Introduction} \label{sec:intro}

Most massive galaxies in the local universe host supermassive black holes (SMBHs).
One fundamental goal in astrophysics is to understand the origins of these SMBHs and their host galaxies.
Galaxy mergers play an important role in their evolution on a cosmological timescale 
\citep[e.g.,][]{Kormendy_&_Ho_2013};
in the assembly of massive galaxies and fueling gas to nuclear SMBHs.
A natural outcome of galaxy mergers containing SMBHs is the formation of binary SMBHs.
If the binary SMBHs coalescence within a Hubble time, a significant fraction of their
rest-mass energy is emitted as gravitational waves (GWs).
The GW emission is a good tool to probe the cosmological evolution of SMBHs
in the framework of hierarchical structure formation.

Pulsar timing array (PTA) experiments enable us to directly address the GW emission in the nHz$-\mu$Hz band.
There are three ongoing PTA experiments:
the European Pulsar Timing Array (EPTA), the Australian Parkes Pulsar Timing Array (PPTA) and 
the North American Nanohertz Observatory for Gravitational Waves (NANOGrav).
Their measurements have recently provided upper limits on the strength of the GW background (GWB)
from binary SMBHs \citep{Arzoumanian_2016,Shannon_2015,Lentati_2015}.

The theoretical GWB signal in the PTA band has been predicted by using semi-analytical calculations
\citep[e.g.,][]{Jaffe_Backer_2003} and cosmological simulations \citep[e.g.,][]{Sesana_2009}.
Those models incorporated various combinations of physical processes that affect
the evolution of binary SMBHs in merging galaxies \citep{BBR_1980};
eccentric binary evolution \citep{Enoki_Nagashima_2007},
viscous drag from a circumbinary gaseous disk \citep{Kocsis_Sesana_2011},
dynamical friction \citep{McWilliams_2014,Kulier_2015}
and multi-body BH interactions \citep{Ryu_2018,Bonetti_2018}.
Combined with cosmological hydrodynamical simulations,
\cite{Kelley_2017} investigated the impact of various environmental processes on the binary evolution.
In spite of great efforts in theory, there are still many uncertainties for model parameters 
and observed empirical relations \citep{Sesana_2013}.

In this {\it Letter}, we address this issue with a different approach following observational results.
We consider ultra-luminous infrared galaxies (ULIRGs), which are one of the best tracers of 
merging galaxies containing SMBHs, as sources of a GWB.
Recent observations by {\it Herschel} and the {\it Wide-field Infrared Survey Explorer} ({\it WISE}) have 
provided a large sample of bright IR galaxies at $0<z<4$  and allow us to explore their infrared 
spectral energy distributions \citep[SEDs;][]{Delvecchio_2014}.
Adopting the luminosity function of ULIRGs, we study the development of a GWB due to 
coalescing binary SMBHs driven by merging ULIRGs.
Intriguingly, we find a tension with the most stringent PTA upper limits, which constrain 
the fraction of binary SMBHs that coalescence by the present time, 
depending on the Eddington ratio of BHs in ULIRGs, the typical lifetime of ULIRGs, and 
the fraction of ULIRGs associated with SMBH binaries.

\vspace{3mm}
\section{Ultra-luminous infrared galaxies hosting SMBHs}\label{sec:ULIRG}

We here consider ULIRGs with total infrared luminosities of $L_{\rm IR,tot}\ga 10^{12}~\lsun$ 
as good tracers of merging galaxies that host at least one accreting SMBH in each system.
At low-redshifts of $z<1$, infrared observations have revealed that the morphologies of
ULIRGs are almost exclusively caused by mergers \citep[e.g.,][]{Surace_1998}, 
and that the number fraction containing AGN increases with $L_{\rm IR,tot}$ and reaches almost unity in ULIRGs
\citep[e.g.,][]{Veilleux_2002,Ichikawa_2014}.
At higher redshifts of $z>1$, the merger fraction is still uncertain since 
active star formation in gas-rich galaxies alone could produce a similar level of 
infrared luminosities observed as ULIRGs \citep[e.g.,][]{Kartaltepe_2012}.

Recent observations have discovered extremely bright ULIRGs with $L_{\rm IR,tot}\simeq 10^{13}-10^{14}~\lsun$, 
the so-called hyper-luminous infrared galaxies \citep[HyLIRGs; ][]{Assef_2015,Tsai_2015}.
This population shows relatively clear signatures 
of galaxy mergers even at higher redshifts.
The merger fraction has been estimated by quantifying their morphologies
as $f_{\rm m,gal}\sim 62\pm 14\%$ for hot dust-obscured galaxies at 
$z\sim 3$ \citep{Fan_2016} and $\sim 75\%$ for HyLIRGs at $1.8<z<2.5$ \citep{Farrah_2017}.
Moreover, the merger fraction tends to increase with bolometric luminosity
and becomes $\sim 80~\%$ at the brightest end of $L_{\rm bol}\sim (1-5)\times 10^{14}~\lsun$ \citep{Glikman_2015}.
Those measurements give a lower limit of the fraction because such merger signatures would smooth out 
after several dynamical timescales.

The enormous power of U/HyLIRGs is produced by deeply buried and rapidly accreting SMBHs.
Most of the AGN radiation is absorbed by surrounding dust and is re-emitted at mid-infrared wavelengths.
By decomposing the hot dust emission from their SEDs, the AGN contribution to the total (IR) luminosity 
increases with $L_{\rm IR,tot}$ \citep[e.g.,][]{Murphy_2011}.
Namely, the luminosity ratio is estimated as $L_{\rm IR, AGN}/L_{\rm IR, tot}\simeq 0.2-0.3$ for ULIRGs 
\citep[e.g.,][]{Ichikawa_2014}, and reaches $\simeq 0.7-1.0$ for HyLIRGs \citep[e.g.,][]{Jones_2014,Farrah_2017}.

\begin{figure*}[htbp]
 \begin{minipage}{0.5\hsize}
   \includegraphics[width=86mm]{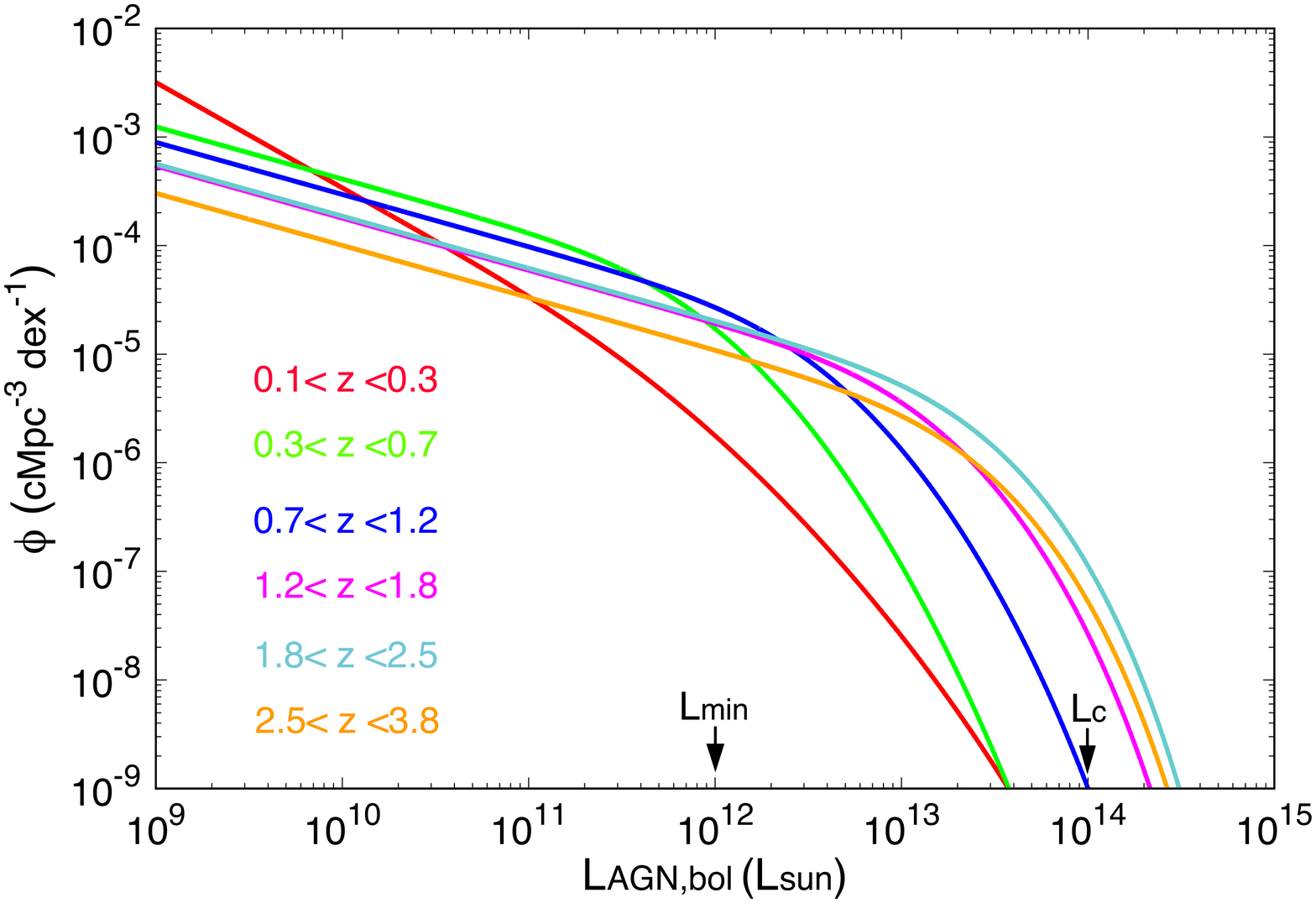}
 \end{minipage}
 \hspace{1mm}
 \begin{minipage}{0.5\hsize}
   \includegraphics[width=86mm]{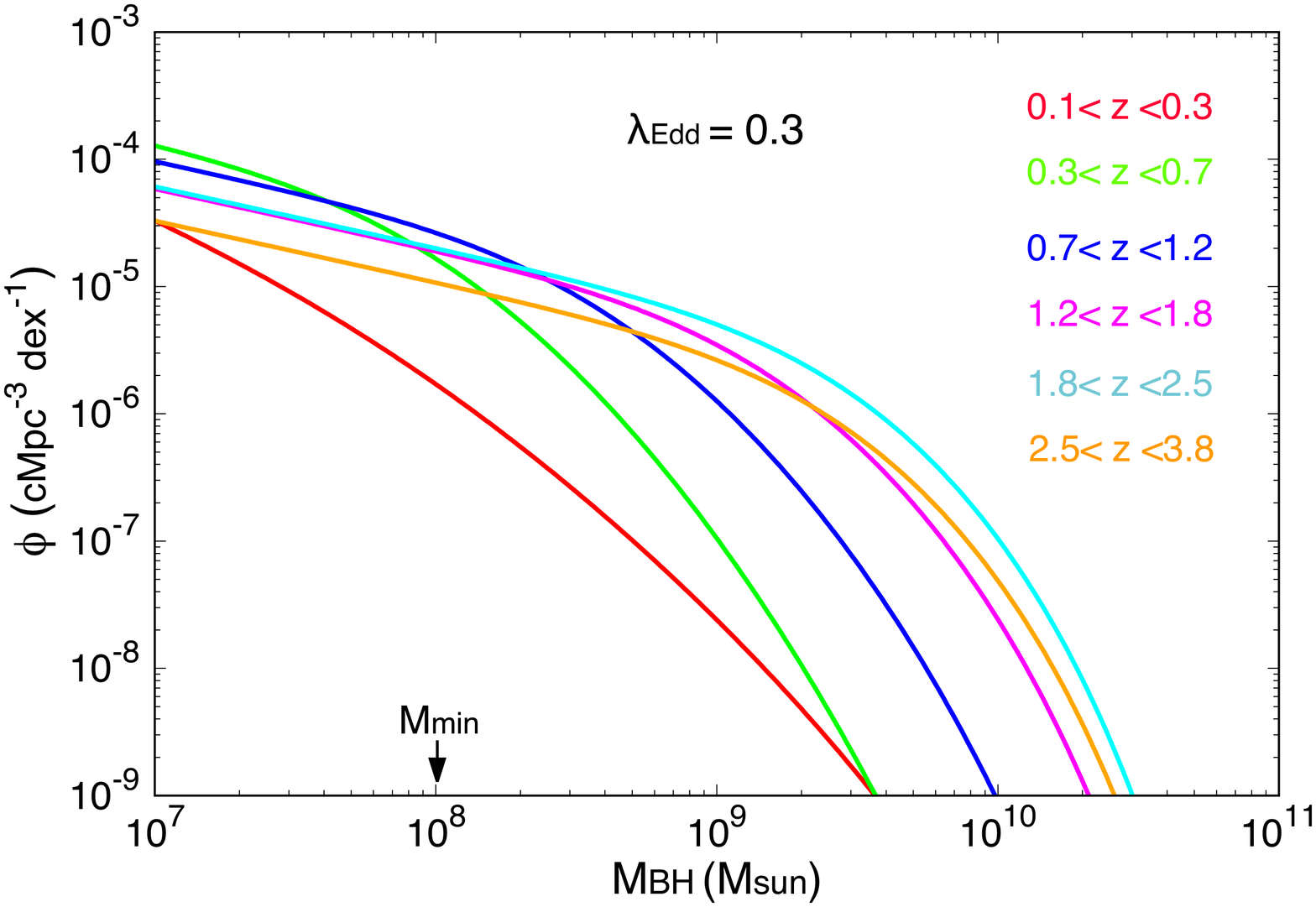}
 \end{minipage}
  \caption{{\it Left}: bolometric luminosity function of AGNs in IR galaxies at different redshifts ($0.1<z<3.8$).
  The functional form is taken from \cite{Delvecchio_2014}, which is based on 
  a sample of Herschel-selected galaxies within the GOODS-S and COSMOS fields.
  {\it Right}: BH mass function inferred from the luminosity function by assuming a constant Eddington ratio
  $\lambda_{\rm Edd}(\equiv L_{\rm bol}/L_{\rm Edd})=0.3$.
 Ultra-luminous IR sources with $L_{\rm bol}\ga 10^{12}~\lsun$ 
 (i.e., $M_{\rm BH}\ga 10^8~\msun$) 
 are considered to be merging galaxies in which coalescing binary SMBHs contribute to the GWB.
}
  \label{fig:LFMF}
\vspace{3mm}
\end{figure*}

Using the SED decomposition technique for a sample of Herschel selected galaxies within the 
Great Observatories Origins Deep Survey-South (GOODS-S) and the Cosmic Evolution Survey (COSMOS) fields, 
\cite{Delvecchio_2014} have reconstructed the AGN bolometric luminosity function (LF) at $0.1 < z < 3.8$.
The bolometric correction factor is estimated by solving the radiative transfer equation for 
a smooth dusty structure irradiated by the AGN accretion disk, instead of adopting the 
bolometric correction shown in \cite{Hopkins_2007a}.
The AGN bolometric LF for ULIRGs is well fit by
\begin{eqnarray}
\frac{d\phi (L,z)}{d\log L } = \phi_\star \left(\frac{L}{L_\star}\right)^{\beta}
\exp \left[-\frac{ \{\log (1+\frac{L}{L_\star})\}^2}{2\sigma^2} -\frac{L}{L_{\rm c}}\right],
\label{eq:LF}
\end{eqnarray}
where $L$ is the AGN bolometric luminosity, and the values of fitting parameters 
($\phi_\star$, $L_\star$, $\beta$ and $\sigma$) are listed in Table 1 of \cite{Delvecchio_2014}.
Here, we set an exponential cutoff above a critical luminosity $L_{\rm c}(\equiv 10^{14}~\lsun)$
because there is no detection for bright AGNs with $L>10^{14}~\lsun$ due to 
the lack of such bright ones or due to the limited observation volume.
In addition, we set a minimum value of the AGN luminosity driven by galaxy major mergers 
to $L_{\rm min}=10^{12}~\lsun$.
Considering $L/L_{\rm IR,AGN}\simeq 3$ \citep{Delvecchio_2014} and 
$L_{\rm IR, AGN}/L_{\rm IR, tot} \simeq 0.2$--$0.3$ \citep{Ichikawa_2014}, 
$L_{\rm min}=10^{12}~\lsun$ corresponds to $L_{\rm IR,tot}\simeq (1-2) \times 10^{12}~\lsun$.
Since the merger fraction of IR galaxies with $L_{\rm IR,tot}< 10^{12}~\lsun$ may be significantly below
unity at high redshift, we do not consider such galaxies as sources of a GWB.
Fig.~\ref{fig:LFMF} (left panel) shows the luminosity function of ULIRGs for different redshifts.

In order to convert the LF to the BH mass function (MF), 
it is necessary to obtain the Eddington ratio ($\lambda_{\rm Edd}=L/L_{\rm Edd}$).
Since the broad line regions of ULIRGs are completely obscured in the optical, 
it is almost impossible to estimate their BH masses and thus the Eddington ratio 
using the optical spectra.
However, a well-defined sample of quasars obtained from the Sloan Digital Sky Surveys (SDSS) catalog
suggests that the typical Eddington ratio for those quasars is
$\simeq 0.3$ \citep{Kollmeier_2006}.
On the other hand, the largest SMBHs have a maximum mass limit at 
$M_{\rm BH,max} \sim {\rm a~few}\times 10^{10}~\msun$,
which is nearly independent of redshift \citep{Netzer_2003,Wu_2015}. 
The radiation luminosity from quasars hosting the most massive BHs is estimated as
$L\simeq 10^{15} \lambda_{\rm Edd} ~\lsun (M_{\rm BH,max} /3\times 10^{10}~\msun)$.
Since this luminosity would be $\ga L_{\rm c}$,
we obtain $ \lambda_{\rm Edd} \ga 0.1 (M_{\rm BH,max} /3\times 10^{10}~\msun)^{-1}$.
Thus, we adopt $ \lambda_{\rm Edd}=0.3$ as our fiducial value.
Fig.~\ref{fig:LFMF} (right panel) shows the BH mass function in ULIRGs for different redshifts.
This BH mass function of ULIRGs is consistent with that of SDSS quasars (QSOs) 
obtained by \cite{Kelly_Shen_2013}, where the BH masses are estimated by using the width 
of the broad emission lines and the AGN continuum luminosity.

Luminous QSOs and ULIRGs are much rarer than normal galaxies,
which is expected since those luminous phases have a lifetime shorter than a Hubble time.
The lifetime is one of the most fundamental quantities for 
estimating the intrinsic number density of those luminous objects.
The QSO lifetime can be observationally constrained by several methods \citep[][and references therein]{Martini_2004}.
Overall, the QSO lifetime lies in the range of $1~{\rm Myr} \la t_{\rm life}\la 100~{\rm Myr}$.
Using galaxy merger simulations, \cite{Hopkins_2006} demonstrated that the lifetime tends to 
decrease with increasing luminosity; namely, $t_{\rm life}\simeq 10-50~{\rm Myr}$ for $L_{\rm bol}>10^{13}~\lsun$.
This shorter lifetime is consistent with $1~{\rm Myr} \la t_{\rm life}\la 20~{\rm Myr}$ obtained from 
observations of extended Ly$\alpha$ emission near luminous QSOs at $2.5\la z \la 2.9$ with ultra-violet 
luminosities of $L_{\rm UV} \sim 10^{14}~\lsun$ \citep{Trainor_Steidel_2013}.
In this letter, we adopt a conservative value of $ t_{\rm life}\simeq 30~{\rm Myr}$ as our fiducial case.

\section{Gravitational wave background}
\label{sec:GWB}

Following \cite{Phinney_2001}, we estimate the GW energy density per logarithmic frequency interval \footnote{
Note that this approach gives the total GW amplitude rather than the unresolved stochastic GWB \citep{Sesana_2008}.
However, individually resolved binaries would contribute at most $10-20~\%$ of the total GW signal in the PTA band. 
Thus, we evaluate the stochastic GWB amplitude using Eq. (\ref{eq:OGW}).}
as
\begin{eqnarray}
\Omega_{\rm GW} (f) = \frac{1}{\rho_{\rm c}c^2}
\int \frac{d^2 N}{d\mathcal{M}_{\rm c} dz} ~\frac{1}{1+z}~
f_r\frac{dE_{\rm gw}}{df_r}~ d\mathcal{M}_{\rm c} dz,
\label{eq:OGW}
\end{eqnarray}
where $\rho_{\rm c}$ is the critical mass density of the Universe at $z=0$,
$f_r$ is the GW frequency in the source's cosmic rest frame, $f=f_r/(1+z)$ is the observed GW frequency,
$z$ is the redshift when the GWs are produced,
$d^2N/d\mathcal{M}_{\rm c}dz$ is the comoving number density of GW events with chirp masses of 
[$\mathcal{M}_{\rm c}, \mathcal{M}_{\rm c}+d\mathcal{M}_{\rm c}$] which occurs at cosmic times 
corresponding to the redshift range between $z$ and $z+dz$,
\begin{eqnarray}
\frac{d^2N}{d\mathcal{M}_{\rm c}dz}~d\mathcal{M}_{\rm c}dz = 
\frac{f_{\rm m,gal}}{f_{\rm duty}}~
\frac{d\phi(L,z)}{d\log L}~d\log L~dz,
\label{eq:dNdMdz}
\end{eqnarray}
where $f_{\rm m,gal}$ is the merger fraction of galaxies inferred from the morphologies of U/HyLIRGs
(we set $f_{\rm m,gal} \simeq 1$; see \S\ref{sec:ULIRG}), and 
$f_{\rm duty} \equiv t_{\rm life}\cdot (dz/dt_r)$ is the duty cycle of ULIRGs.
Since a constant $t_{\rm life}\simeq 30~{\rm Myr}$ is adopted, the duty cycle is independent of the luminosity.

The GW emission spectrum from a merging binary in the rest frame is given by
\begin{equation}
\frac{dE_{\rm gw}}{df_r}=\frac{(\pi G)^{2/3}\mathcal{M}_{\rm c}^{5/3}}{3f_r^{1/3}},
\label{eq:Egw}
\end{equation}
where $E_{\rm gw}$ is the energy of the GW.
The chirp mass is written as $\mathcal{M}_{\rm c}^{5/3}\equiv qM_{\rm BH}^{5/3}/(1+q)^{1/3}$,
where $M_{\rm BH}$ is the mass of the primary SMBH and 
$q(<1)$ is the mass ratio of two SMBHs.
We suppose that the primary BH follows the MF in Figure~\ref{fig:LFMF}.
Thus, we implicitly assume that the primary SMBH is located at the center of a ULIRG after the galaxy mergers
and is responsible for the ULIRG activity in a lifetime of $t_{\rm life}$, 
while the secondary BH is still located off center in a lower-density region.
Since the secondary BH would decay its orbit via dynamical friction (DF) 
on a timescale of $t_{\rm DF}\sim 100$ Myr \citep[e.g.,][]{Yu_2002}, 
the binary formation would occur after the ULIRG phase (i.e., $t_{\rm DF}\ga t_{\rm life}$).
This assumption would be plausible because the number fraction of AGNs that are dual SMBHs is 
as small as $\sim 10~\%$ at $z<1$ \citep{Comerford_2013}.

As discussed in \S\ref{sec:ULIRG}, most ULIRGs are triggered by major mergers of galaxies.
We here set a minimum value of the BH mass ratio to $q_{\rm min}\simeq 0.1$.
The mass-ratio distribution of SMBHs, $\Phi(q)$ at $q_{\rm min}\leq q \leq 1$, is uncertain.
However, the chirp mass averaged over $q$ is less uncertain, namely,
$\langle \mathcal{M}_{\rm c}^{5/3}\rangle / M_{\rm BH}^{5/3} \simeq 0.47$, $0.34$ and $0.3$ 
for $\Phi(q)={\rm const.}$, $\Phi(q)\propto 1/q$ and $\Phi(q)\propto \delta(q-1/3)$, respectively.
We adopt the last one for a more conservative estimate.

\begin{deluxetable}{cccc}[t]
\tablecaption{The relation between the delay time, coalescence fraction and
GWB amplitude.\label{tab:mathmode}}
\tablecolumns{3}
\tablenum{1}
\tablewidth{00pt}
\tablehead{
~~$t_{\rm delay}$(Gyr)~~&
~~$\mathcal{F}_{\rm coal}$~~&
~~$\mathcal{F}_{\rm GW}$~~&
~~$h_c\times 10^{15}$~~ 
}
\vspace{0mm}
\startdata
0 & 1.0 & 1.0 & 3.32\\
3 & 1.0 & 1.16 & 3.57\\
7 & 0.75 & 1.22 & 3.67\\
10 & 0.31 & 0.83 & 3.03\\
11 & 0.14 & 0.39 & 2.06\\
12 & 0.022 & 0.06 & 0.81
\enddata
\tablecomments{(1) delay time, (2) coalescence fraction of BHs,
(3) effective coalescence fraction, (4) total GW amplitude and 
(5) stochastic GWB amplitude at $f=10^{-8}~\hz$.
We adopt $\lambda_{\rm Edd}=0.3$ and $t_{\rm life}=30~{\rm Myr}$.
}
\label{tab1}
\vspace{-19pt}
\end{deluxetable}

In the course of a galaxy merger embedding two SMBHs, 
a variety of physical processes affect the binary evolution
to the coalescence in a certain timescale \citep[e.g.,][]{BBR_1980,Merritt_2013}.
Since the delay-time between formation and coalescence of BH binaries is still uncertain,
rather than attempting to model this delay, we assume a uniform delay time of $t_{\rm delay}$.
To consider the delay effect on BH mergers, we evaluate the LF and duty cycle in Eq. (\ref{eq:dNdMdz}) at 
$\tilde{z}\equiv z+\Delta z$, where $\Delta z$ is determined by solving 
$t_{\rm delay}=\int^z_{z+\Delta z}\frac{dt_r}{dz}dz$.
The delay time tends to be shorter than a Hubble timescale for SMBH binaries with
higher total masses ($M_{\rm BH}>10^8~\msun$ ) and mass ratios ($q>0.2$) \citep{Khan_2016,Kelley_2017}.
Within their model uncertainties, the delay time is estimated as $\simeq 0.35-6.9~{\rm Gyr}$
(see Table B1 in \citealt{Kelley_2017}),
which is significantly longer than the SMBH binary formation ($t_{\rm DF}\sim 100$ Myr) 
and the ULIRG's lifetime ($t_{\rm life}<100$ Myr).

\begin{figure}
\includegraphics[width=1.\linewidth]{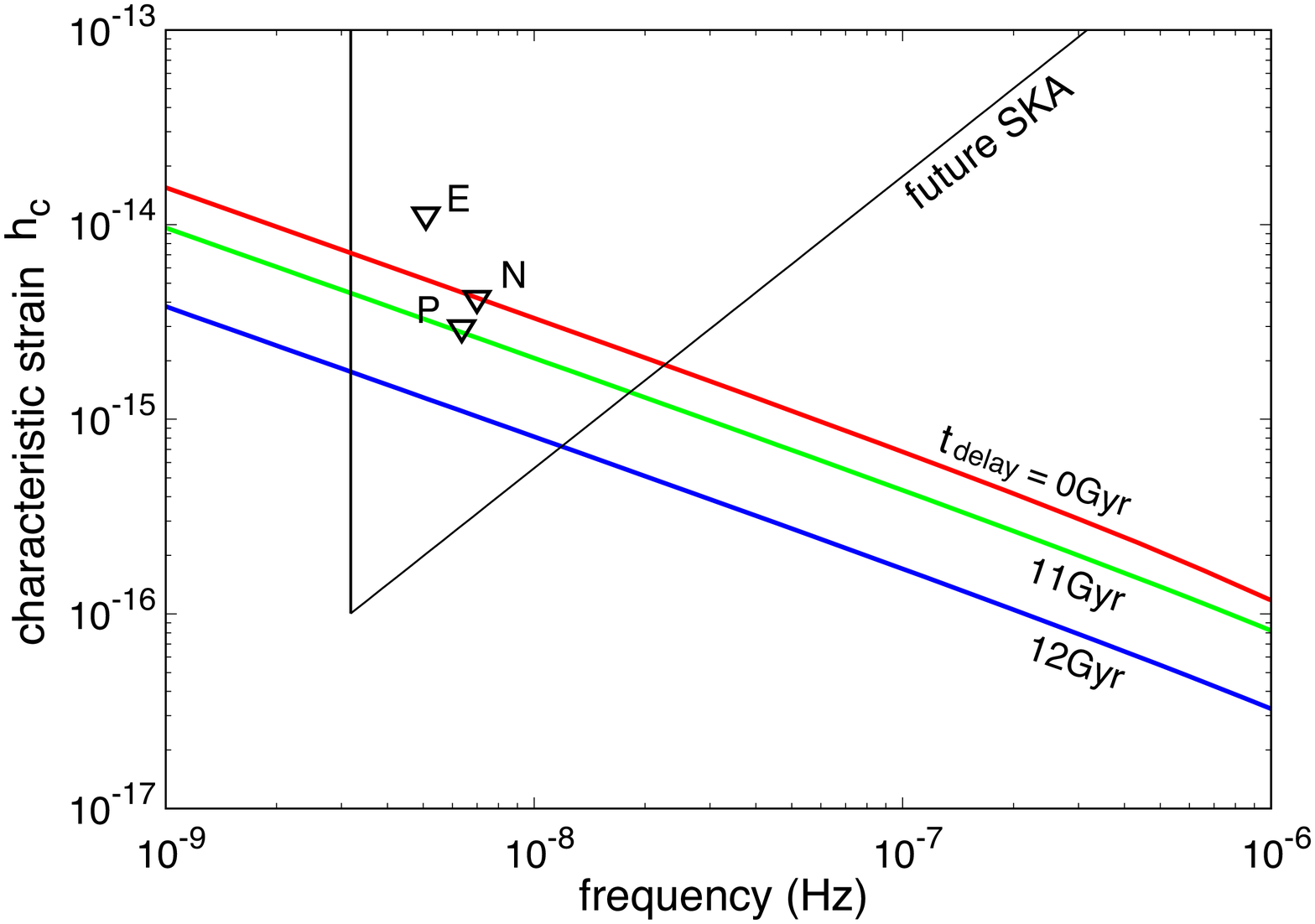}
\caption{Characteristic amplitude of the GWB signal
for different delay timescales for BH mergers ($0\leq t_{\rm delay}\leq 12$ Gyr).
Solid curves are the predicted total GW amplitudes.
Triangle symbols show the current upper limits from PTA experiments:
the EPTA (E), NANOGrav (N), and PPTA (P).
To be consistent with the PTA limits, $t_{\rm delay} \ga 11$ Gyr is required.
Black solid line refers to the expected sensitivity by the SKA assuming monitoring of 50 pulsars 
at 100 ns rms precision over $T_{\rm obs}=10$ yr with a cadence of 20 yr$^{-1}$.
}
\label{fig:h_c}
\vspace{2mm}
\end{figure}

By using Eqs. (\ref{eq:OGW}), (\ref{eq:dNdMdz}), and (\ref{eq:Egw}), 
the energy spectrum of the total GW emission due to merging SMBHs 
in ULIRGs is calculated as 
\begin{eqnarray}
\Omega_{\rm GW}(f) &=& 1.53\times 10^{-9}~
\mathcal{F}_{\rm GW}\\
& \times &
\left(\frac{\lambda_{\rm Edd}}{0.3}\right)^{-5/3}
\left(\frac{t_{\rm life}}{30~{\rm Myr}}\right)^{-1}
\left(\frac{f}{10~{\rm nHz}}\right)^{2/3}, \nonumber
\end{eqnarray}
and the characteristic strain is estimated as 
\begin{eqnarray}
h_{\rm c}(f) &=& 3.32\times 10^{-15}~
\mathcal{F}_{\rm GW}^{1/2} \\
&\times &
\left(\frac{\lambda_{\rm Edd}}{0.3}\right)^{-5/6}
\left(\frac{t_{\rm life}}{30~{\rm Myr}}\right)^{-1/2}
\left(\frac{f}{10~{\rm nHz}}\right)^{-2/3},\nonumber
\end{eqnarray}
where $\mathcal{F}_{\rm GW}$ is the ratio of the GW amplitude 
with an assumed value of $t_{\rm delay}$ to that without delay.
We also define the number fraction of SMBHs that coalesce within a Hubble time
as $\mathcal{F}_{\rm coal}$.
As shown in Table~\ref{tab1}, the coalescence fraction decreases monotonically with $t_{\rm delay}$
and drops sharply at $t_{\rm delay}\ga 7$ Gyr.
On the other hand, the GW amplitude slightly increases with $t_{\rm delay}$
(i.e., $\mathcal{F}_{\rm GW}>1$), and decreases at $t_{\rm delay}\ga 10$ Gyr
significantly (i.e., $\mathcal{F}_{\rm GW}<1$).
This is because a short delay time barely reduces the number of merger events occurring in a Hubble time, 
but induces BH mergers at lower redshift.
In Figure \ref{fig:h_c}, we plot the total GW spectrum of interest (solid).
The upper limits from the PTA are presented by triangle symbols at frequencies 
where the limit becomes the most stringent; 
EPA \citep{Lentati_2015}, NANOGrav \citep{Arzoumanian_2016} and PPTA 
\citep{Shannon_2015}.
Without the delay effect, merging SMBHs in ULIRGs would overproduce a GWB.
The GWB has also been overproduced based on the observed \citep{McWilliams_2014} 
and simulated \citep{Kulier_2015} abundance of massive galaxies, and based on 
the observed periodic quasar candidates \citep{Sesana_2018},
assuming that these objects all host SMBH binary mergers. 
With these PTA constraints, we obtain an upper limit for $\mathcal{F}_{\rm GW}$ as
\begin{eqnarray}
\mathcal{F}_{\rm GW} \la 0.43
\left(\frac{\lambda_{\rm Edd}}{0.3}\right)^{5/3}
\left(\frac{t_{\rm life}}{30~{\rm Myr}}\right),
\label{eq:result}
\end{eqnarray}
which corresponds to $t_{\rm delay}\ga 10.9$ Gyr and $\mathcal{F}_{\rm coal}\la 0.16$.
We also plot the sensitivity that is achievable with the Square Kilometer Array (SKA) with 
50 pulsars for a $T_{\rm obs}=10$ yr observation.
If a GWB will not be detected even by such a planned detector, it would imply a strong constraint 
on $\mathcal{F}_{\rm coal} \ll 1\%$.

\begin{figure}
\includegraphics[width=1.\linewidth]{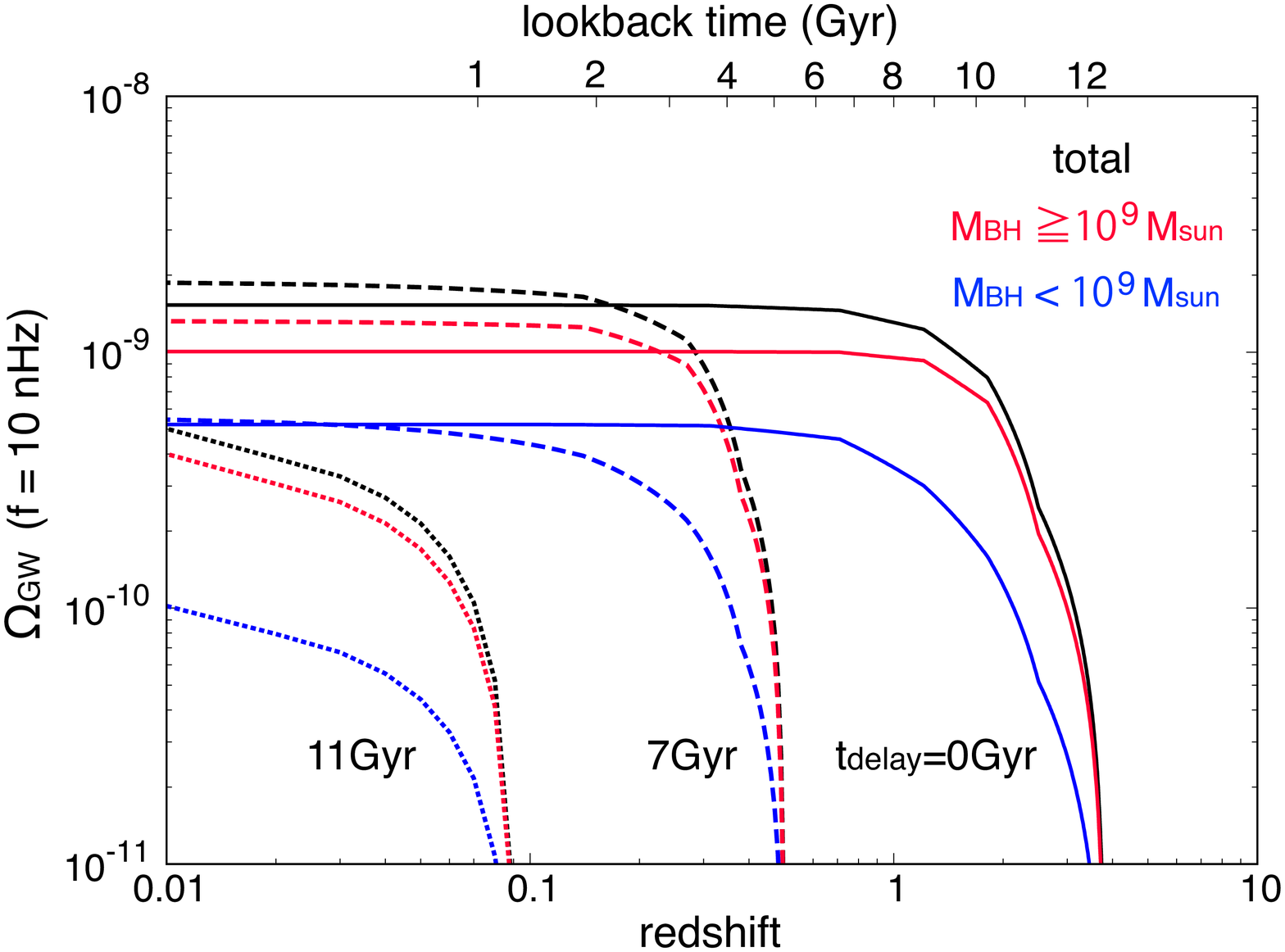}
\caption{Evolution of the stochastic GWB produced by coalescing SMBHs in ULIRGs
without and with the delay of BH mergers; $t_{\rm delay}=0$, $7$ and $11$ Gyr.
Black curves show the total GWB, and others present the contribution from SMBHs 
with masses of $M_{\rm BH}\geq 10^{9}~\msun$ (red) and $<10^9~\msun$ (blue).
}
\label{fig:Ogw}
\vspace{3mm}
\end{figure}

In Figure \ref{fig:Ogw}, we present the evolution of the stochastic 
GWB (black), and those due to SMBHs with $M_{\rm BH}\geq 10^{9}~\msun$ (red) and $<10^9~\msun$ (blue).
Without the delay, half of the stochastic 
GWB energy is produced by merging SMBHs with 
$M_{\rm BH}\ga 10^9~\msun$ at $z>1.5$, while others are due to less massive ones at $z<1.5$.
This result reflects the shape and redshift-evolution of the LF of ULIRGs and MF of SMBHs.
In fact, a GWB from higher-mass BHs dominate at higher redshift.  
The GWB amplitude decreases significantly at $t_{\rm delay}\ga 10$ Gyr because 
a larger fraction of SMBHs in ULIRGs do not merge within a Hubble time.
Our results are {\it qualitatively} consistent with previous work 
\cite[e.g.,][]{Sesana_2008}, concluding that a GBW in the PTA band 
is dominated by nearby and massive binary SMBHs ($z<2$ and $M_{\rm BH}>10^8~\msun$).
However, it is worth emphasizing that coalescing binary SMBHs even in 
a rare population of high-z ULIRGs associated gas-rich major mergers,
which are quite different from multiple dry mergers occurring at low redshift,
can produce a GWB close to the present-day upper limit.
Since other type of galaxies unlike ULIRGs may have additional SMBH mergers 
and contribute to the GWBs \citep[see e.g.,][]{McWilliams_2014},
our result gives a lower limit on the total GWB in the PTA band.

We briefly discuss possible biased estimates of the BH masses caused by a scatter of 
the Eddington ratio distribution \citep{Shen_2008}.
For the SDSS quasar sample, the Eddington ratio approximately follows a log-normal distribution 
with a dispersion of $\sigma_{\rm E}\simeq 0.25$ dex for a fixed BH mass \citep[see Figure 12 in][]{Kelly_Shen_2013}.
We estimate the mass bias as $\gamma_{\rm L} \sigma_{\rm E}^2 \ln 10$ dex, 
assuming a power-law shape for the underlying true mass function and a symmetric Gaussian scatter in 
$\log M_{\rm BH}$ around a mass-independent mean value, where $\gamma_{\rm L}(<0)$ is the slope 
of the AGN bolometric LF, and find that the bias effect reduces the GWB amplitude by half.
However, we note that the mean Eddington ratio for SMBHs of interest is significantly lower than our fiducial value \citep{Kelly_Shen_2013}.
Therefore, the GWB amplitude would rather be enhanced ($\Omega_{\rm GW}\propto \lambda_{\rm Edd}^{-5/3}$), and 
the bias effect would be cancelled out.

\vspace{3mm}
\section{Discussion and Implications}\label{sec:dis}

The growth of SMBHs in galactic nuclei can be constrained by observations of present-day BH remnants.
Adopting the LF of ULIRGs and a radiative efficiency $\epsilon_r$, 
we can estimate the BH mass density accreted during the ULIRG phases, 
and compare it to that observed in the local universe \citep{Soltan_1982,Yu_Tremaine_2002}.
The estimated BH mass density is given by
\begin{eqnarray}
\rho_{\rm BH} \simeq 3.5 \times 10^4~\left(\frac{1-\epsilon_r}{\epsilon_r}\right)~\msun~\cmpc^{-3},
\end{eqnarray}
\citep{Delvecchio_2014}.
In order not to exceed the value observed in the local universe,
$\rho_{\rm BH,obs} \simeq 4.2^{+1.2}_{-1.0} \times 10^5~\msun~\cmpc^{-3}$ \citep[e.g.,][]{Shankar_2009},
the radiative efficiency is required to be $\epsilon_r \ga 0.076^{+0.023}_{-0.018}$.
This efficiency is consistent with similar arguments for bright QSOs in the optical and X-ray bands
\cite[e.g.,][]{Yu_Tremaine_2002,Hopkins_2007a,Ueda_2014}.

The total present-day energy density in GW radiation 
is estimated from
\begin{eqnarray}
& \mathcal{E}_{\rm GW}& \equiv \int ^{\infty}_0 \rho_{\rm c}c^2\Omega_{\rm GW}(f)\frac{df}{f},
\label{eq:total_EGW}
\end{eqnarray}
and $\mathcal{E}_{\rm GW}\simeq 4.3\times 10^{-16}~{\rm erg~s}^{-1}~\cc$
for $t_{\rm delay}=11$ Gyr, $\lambda_{\rm Edd}=0.3$ and $t_{\rm life}=30$ Myr.
Note that $\mathcal{E}_{\rm GW} \propto \lambda_{\rm Edd}^{-0.98}t_{\rm life}^{-1}$.
As a result, the ratio of the total GW energy to the present-day SMBH rest 
mass energy is estimated as 
\begin{eqnarray}
\frac{\mathcal{E}_{\rm GW}}{\rho_{\rm BH}c^2} 
\la \frac{0.20~\epsilon_r}{1-\epsilon_r},
\end{eqnarray}
because $t_{\rm delay}\ga 11$ Gyr is required from the PTA observations.
We note that for unequal-mass binaries, the GW radiative efficiency from a BH with a mass of $qM$,
which is gravitationally captured in a circular orbit by a BH with mass of $M$, is given by
$\epsilon _{gw}\simeq (0.057+0.4448\eta + 0.522\eta^2)/(1+q)\simeq 0.1192$, where $\eta=q/(1+q)^2$
and $q=1/3$ is set \citep{Lousto_2010}.
Approximating $\epsilon _{gw}\simeq \epsilon _{r}$, therefore we obtain the interesting constraint that
the contribution of BH mergers to the present-day BH mass density is less than 
$\la 20 \langle 1+z\rangle \%$; see Eq. 7 in \cite{Phinney_2001}.

The brightest U/HyLIRGs that have experienced active star formation at high redshift
would be observed as massive elliptical galaxies 
in the local universe.
An important consequence from Eq. (\ref{eq:result}) is that 
$\ga 80\%$ of the binary SMBHs formed in ULIRGs neither coalesce within a Hubble time 
nor contribute to the GWB.
For SMBH binaries with mass ratios of $q>0.1$, the dynamical friction caused by
surrounding stars with velocity dispersion $\sigma_\star$ would carry the binary separation down to
\begin{eqnarray}
a_{\rm h} \sim \frac{12q}{1+q}
\left(\frac{M_{\rm BH}}{10^9~\msun}\right)
\left(\frac{\sigma_\star}{300~\kms}\right)^{-2}~\pc,
\end{eqnarray}
where the binary becomes hard, ejects stars and stalls the orbital decay \citep[e.g.,][]{Merritt_2013}.
The formation timescale of an SMBH hard binary is typically $\sim 100~{\rm Myr}$ \citep[e.g.,][]{Yu_2002},
which is much shorter than both a Hubble time and the delay time $t_{\rm delay}$ of interest. 
This suggests that a remnant population of $O(1-10)$ pc binaries would be left 
at the centers of nearby massive ellipticals.
Although no such binaries have been detected by PTAs to date, this non-detection has already yielded interesting constraints
on their mass ratios \citep{Schutz_Ma_2016} and anisotropy in the GWB \citep{Mingarelli_2017}.

\acknowledgments
We thank Jeremiah Ostriker, Alberto Sesana and Yoshiki Toba for useful discussions and comments.
This work is partially supported by the Simons Foundation through the Simons Society of Fellows (KI),
by JSPS KAKENHI (18K13584; KI), and by NASA grant NNX17AL82G and NSF grant 1715661 (ZH).


\begin{thebibliography}{}
\expandafter\ifx\csname natexlab\endcsname\relax\def\natexlab#1{#1}\fi
\providecommand{\url}[1]{\href{#1}{#1}}
\providecommand{\dodoi}[1]{doi:~\href{http://doi.org/#1}{\nolinkurl{#1}}}
\providecommand{\doeprint}[1]{\href{http://ascl.net/#1}{\nolinkurl{http://ascl.net/#1}}}
\providecommand{\doarXiv}[1]{\href{https://arxiv.org/abs/#1}{\nolinkurl{https://arxiv.org/abs/#1}}}

\bibitem[{{Arzoumanian} {et~al.}(2016){Arzoumanian}, {Brazier},
  {Burke-Spolaor}, {Chamberlin}, {Chatterjee}, {Christy}, {Cordes}, {Cornish},
  {Crowter}, {Demorest}, {Deng}, {Dolch}, {Ellis}, {Ferdman}, {Fonseca},
  {Garver-Daniels}, {Gonzalez}, {Jenet}, {Jones}, {Jones}, {Kaspi}, {Koop},
  {Lam}, {Lazio}, {Levin}, {Lommen}, {Lorimer}, {Luo}, {Lynch}, {Madison},
  {McLaughlin}, {McWilliams}, {Mingarelli}, {Nice}, {Palliyaguru}, {Pennucci},
  {Ransom}, {Sampson}, {Sanidas}, {Sesana}, {Siemens}, {Simon}, {Stairs},
  {Stinebring}, {Stovall}, {Swiggum}, {Taylor}, {Vallisneri}, {van Haasteren},
  {Wang}, {Zhu}, \& {NANOGrav Collaboration}}]{Arzoumanian_2016}
{Arzoumanian}, Z., {Brazier}, A., {Burke-Spolaor}, S., {et~al.} 2016, \apj,
  821, 13, \dodoi{10.3847/0004-637X/821/1/13}

\bibitem[{{Assef} {et~al.}(2015){Assef}, {Eisenhardt}, {Stern}, {Tsai}, {Wu},
  {Wylezalek}, {Blain}, {Bridge}, {Donoso}, {Gonzales}, {Griffith}, \&
  {Jarrett}}]{Assef_2015}
{Assef}, R.~J., {Eisenhardt}, P.~R.~M., {Stern}, D., {et~al.} 2015, \apj, 804,
  27, \dodoi{10.1088/0004-637X/804/1/27}

\bibitem[{{Begelman} {et~al.}(1980){Begelman}, {Blandford}, \&
  {Rees}}]{BBR_1980}
{Begelman}, M.~C., {Blandford}, R.~D., \& {Rees}, M.~J. 1980, \nat, 287, 307,
  \dodoi{10.1038/287307a0}

\bibitem[{{Bonetti} {et~al.}(2018){Bonetti}, {Haardt}, {Sesana}, \&
  {Barausse}}]{Bonetti_2018}
{Bonetti}, M., {Haardt}, F., {Sesana}, A., \& {Barausse}, E. 2018, \mnras,
  \dodoi{10.1093/mnras/sty896}

\bibitem[{{Comerford} {et~al.}(2013){Comerford}, {Schluns}, {Greene}, \&
  {Cool}}]{Comerford_2013}
{Comerford}, J.~M., {Schluns}, K., {Greene}, J.~E., \& {Cool}, R.~J. 2013,
  \apj, 777, 64, \dodoi{10.1088/0004-637X/777/1/64}

\bibitem[{{Delvecchio} {et~al.}(2014){Delvecchio}, {Gruppioni}, {Pozzi},
  {Berta}, {Zamorani}, {Cimatti}, {Lutz}, {Scott}, {Vignali}, {Cresci},
  {Feltre}, {Cooray}, {Vaccari}, {Fritz}, {Le Floc'h}, {Magnelli}, {Popesso},
  {Oliver}, {Bock}, {Carollo}, {Contini}, {Le F{\'e}vre}, {Lilly}, {Mainieri},
  {Renzini}, \& {Scodeggio}}]{Delvecchio_2014}
{Delvecchio}, I., {Gruppioni}, C., {Pozzi}, F., {et~al.} 2014, \mnras, 439,
  2736, \dodoi{10.1093/mnras/stu130}

\bibitem[{{Enoki} \& {Nagashima}(2007)}]{Enoki_Nagashima_2007}
{Enoki}, M., \& {Nagashima}, M. 2007, Progress of Theoretical Physics, 117,
  241, \dodoi{10.1143/PTP.117.241}

\bibitem[{{Fan} {et~al.}(2016){Fan}, {Han}, {Fang}, {Gao}, {Zhang}, {Jiang},
  {Wu}, {Yang}, \& {Li}}]{Fan_2016}
{Fan}, L., {Han}, Y., {Fang}, G., {et~al.} 2016, \apjl, 822, L32,
  \dodoi{10.3847/2041-8205/822/2/L32}

\bibitem[{{Farrah} {et~al.}(2017){Farrah}, {Petty}, {Connolly}, {Blain},
  {Efstathiou}, {Lacy}, {Stern}, {Lake}, {Jarrett}, {Bridge}, {Eisenhardt},
  {Benford}, {Jones}, {Tsai}, {Assef}, {Wu}, \& {Moustakas}}]{Farrah_2017}
{Farrah}, D., {Petty}, S., {Connolly}, B., {et~al.} 2017, \apj, 844, 106,
  \dodoi{10.3847/1538-4357/aa78f2}

\bibitem[{{Glikman} {et~al.}(2015){Glikman}, {Simmons}, {Mailly}, {Schawinski},
  {Urry}, \& {Lacy}}]{Glikman_2015}
{Glikman}, E., {Simmons}, B., {Mailly}, M., {et~al.} 2015, \apj, 806, 218,
  \dodoi{10.1088/0004-637X/806/2/218}

\bibitem[{{Hopkins} {et~al.}(2006){Hopkins}, {Hernquist}, {Cox}, {Di Matteo},
  {Robertson}, \& {Springel}}]{Hopkins_2006}
{Hopkins}, P.~F., {Hernquist}, L., {Cox}, T.~J., {et~al.} 2006, \apjs, 163, 1,
  \dodoi{10.1086/499298}

\bibitem[{{Hopkins} {et~al.}(2007){Hopkins}, {Richards}, \&
  {Hernquist}}]{Hopkins_2007a}
{Hopkins}, P.~F., {Richards}, G.~T., \& {Hernquist}, L. 2007, \apj, 654, 731,
  \dodoi{10.1086/509629}

\bibitem[{{Ichikawa} {et~al.}(2014){Ichikawa}, {Imanishi}, {Ueda}, {Nakagawa},
  {Shirahata}, {Kaneda}, \& {Oyabu}}]{Ichikawa_2014}
{Ichikawa}, K., {Imanishi}, M., {Ueda}, Y., {et~al.} 2014, \apj, 794, 139,
  \dodoi{10.1088/0004-637X/794/2/139}

\bibitem[{{Jaffe} \& {Backer}(2003)}]{Jaffe_Backer_2003}
{Jaffe}, A.~H., \& {Backer}, D.~C. 2003, \apj, 583, 616, \dodoi{10.1086/345443}

\bibitem[{{Jones} {et~al.}(2014){Jones}, {Blain}, {Stern}, {Assef}, {Bridge},
  {Eisenhardt}, {Petty}, {Wu}, {Tsai}, {Cutri}, {Wright}, \&
  {Yan}}]{Jones_2014}
{Jones}, S.~F., {Blain}, A.~W., {Stern}, D., {et~al.} 2014, \mnras, 443, 146,
  \dodoi{10.1093/mnras/stu1157}

\bibitem[{{Kartaltepe} {et~al.}(2012){Kartaltepe}, {Dickinson}, {Alexander},
  {Bell}, {Dahlen}, {Elbaz}, {Faber}, {Lotz}, {McIntosh}, {Wiklind}, {Altieri},
  {Aussel}, {Bethermin}, {Bournaud}, {Charmandaris}, {Conselice}, {Cooray},
  {Dannerbauer}, {Dav{\'e}}, {Dunlop}, {Dekel}, {Ferguson}, {Grogin}, {Hwang},
  {Ivison}, {Kocevski}, {Koekemoer}, {Koo}, {Lai}, {Leiton}, {Lucas}, {Lutz},
  {Magdis}, {Magnelli}, {Morrison}, {Mozena}, {Mullaney}, {Newman}, {Pope},
  {Popesso}, {van der Wel}, {Weiner}, \& {Wuyts}}]{Kartaltepe_2012}
{Kartaltepe}, J.~S., {Dickinson}, M., {Alexander}, D.~M., {et~al.} 2012, \apj,
  757, 23, \dodoi{10.1088/0004-637X/757/1/23}

\bibitem[{{Kelley} {et~al.}(2017){Kelley}, {Blecha}, \&
  {Hernquist}}]{Kelley_2017}
{Kelley}, L.~Z., {Blecha}, L., \& {Hernquist}, L. 2017, \mnras, 464, 3131,
  \dodoi{10.1093/mnras/stw2452}

\bibitem[{{Kelly} \& {Shen}(2013)}]{Kelly_Shen_2013}
{Kelly}, B.~C., \& {Shen}, Y. 2013, \apj, 764, 45,
  \dodoi{10.1088/0004-637X/764/1/45}

\bibitem[{{Khan} {et~al.}(2016){Khan}, {Fiacconi}, {Mayer}, {Berczik}, \&
  {Just}}]{Khan_2016}
{Khan}, F.~M., {Fiacconi}, D., {Mayer}, L., {Berczik}, P., \& {Just}, A. 2016,
  \apj, 828, 73, \dodoi{10.3847/0004-637X/828/2/73}

\bibitem[{{Kocsis} \& {Sesana}(2011)}]{Kocsis_Sesana_2011}
{Kocsis}, B., \& {Sesana}, A. 2011, \mnras, 411, 1467,
  \dodoi{10.1111/j.1365-2966.2010.17782.x}

\bibitem[{{Kollmeier} {et~al.}(2006){Kollmeier}, {Onken}, {Kochanek}, {Gould},
  {Weinberg}, {Dietrich}, {Cool}, {Dey}, {Eisenstein}, {Jannuzi}, {Le Floc'h},
  \& {Stern}}]{Kollmeier_2006}
{Kollmeier}, J.~A., {Onken}, C.~A., {Kochanek}, C.~S., {et~al.} 2006, \apj,
  648, 128, \dodoi{10.1086/505646}

\bibitem[{{Kormendy} \& {Ho}(2013)}]{Kormendy_&_Ho_2013}
{Kormendy}, J., \& {Ho}, L.~C. 2013, \araa, 51, 511,
  \dodoi{10.1146/annurev-astro-082708-101811}

\bibitem[{{Kulier} {et~al.}(2015){Kulier}, {Ostriker}, {Natarajan}, {Lackner},
  \& {Cen}}]{Kulier_2015}
{Kulier}, A., {Ostriker}, J.~P., {Natarajan}, P., {Lackner}, C.~N., \& {Cen},
  R. 2015, \apj, 799, 178, \dodoi{10.1088/0004-637X/799/2/178}

\bibitem[{{Lentati} {et~al.}(2015){Lentati}, {Taylor}, {Mingarelli}, {Sesana},
  {Sanidas}, {Vecchio}, {Caballero}, {Lee}, {van Haasteren}, {Babak}, {Bassa},
  {Brem}, {Burgay}, {Champion}, {Cognard}, {Desvignes}, {Gair}, {Guillemot},
  {Hessels}, {Janssen}, {Karuppusamy}, {Kramer}, {Lassus}, {Lazarus}, {Liu},
  {Os{\l}owski}, {Perrodin}, {Petiteau}, {Possenti}, {Purver}, {Rosado},
  {Smits}, {Stappers}, {Theureau}, {Tiburzi}, \& {Verbiest}}]{Lentati_2015}
{Lentati}, L., {Taylor}, S.~R., {Mingarelli}, C.~M.~F., {et~al.} 2015, \mnras,
  453, 2576, \dodoi{10.1093/mnras/stv1538}

\bibitem[{{Lousto} {et~al.}(2010){Lousto}, {Campanelli}, {Zlochower}, \&
  {Nakano}}]{Lousto_2010}
{Lousto}, C.~O., {Campanelli}, M., {Zlochower}, Y., \& {Nakano}, H. 2010,
  Classical and Quantum Gravity, 27, 114006,
  \dodoi{10.1088/0264-9381/27/11/114006}

\bibitem[{{Martini}(2004)}]{Martini_2004}
{Martini}, P. 2004, Coevolution of Black Holes and Galaxies, 169

\bibitem[{{McWilliams} {et~al.}(2014){McWilliams}, {Ostriker}, \&
  {Pretorius}}]{McWilliams_2014}
{McWilliams}, S.~T., {Ostriker}, J.~P., \& {Pretorius}, F. 2014, \apj, 789,
  156, \dodoi{10.1088/0004-637X/789/2/156}

\bibitem[{{Merritt}(2013)}]{Merritt_2013}
{Merritt}, D. 2013, {Dynamics and Evolution of Galactic Nuclei}

\bibitem[{{Mingarelli} {et~al.}(2017){Mingarelli}, {Lazio}, {Sesana}, {Greene},
  {Ellis}, {Ma}, {Croft}, {Burke-Spolaor}, \& {Taylor}}]{Mingarelli_2017}
{Mingarelli}, C.~M.~F., {Lazio}, T.~J.~W., {Sesana}, A., {et~al.} 2017, Nature
  Astronomy, 1, 886, \dodoi{10.1038/s41550-017-0299-6}

\bibitem[{{Murphy} {et~al.}(2011){Murphy}, {Chary}, {Dickinson}, {Pope},
  {Frayer}, \& {Lin}}]{Murphy_2011}
{Murphy}, E.~J., {Chary}, R.-R., {Dickinson}, M., {et~al.} 2011, \apj, 732,
  126, \dodoi{10.1088/0004-637X/732/2/126}

\bibitem[{{Netzer}(2003)}]{Netzer_2003}
{Netzer}, H. 2003, \apjl, 583, L5, \dodoi{10.1086/368012}

\bibitem[{{Phinney}(2001)}]{Phinney_2001}
{Phinney}, E.~S. 2001, arXiv:0108028

\bibitem[{{Ryu} {et~al.}(2018){Ryu}, {Perna}, {Haiman}, {Ostriker}, \&
  {Stone}}]{Ryu_2018}
{Ryu}, T., {Perna}, R., {Haiman}, Z., {Ostriker}, J.~P., \& {Stone}, N.~C.
  2018, \mnras, 473, 3410, \dodoi{10.1093/mnras/stx2524}

\bibitem[{{Schutz} \& {Ma}(2016)}]{Schutz_Ma_2016}
{Schutz}, K., \& {Ma}, C.-P. 2016, \mnras, 459, 1737,
  \dodoi{10.1093/mnras/stw768}

\bibitem[{{Sesana}(2013)}]{Sesana_2013}
{Sesana}, A. 2013, \mnras, 433, L1, \dodoi{10.1093/mnrasl/slt034}

\bibitem[{{Sesana} {et~al.}(2018){Sesana}, {Haiman}, {Kocsis}, \&
  {Kelley}}]{Sesana_2018}
{Sesana}, A., {Haiman}, Z., {Kocsis}, B., \& {Kelley}, L.~Z. 2018, \apj, 856,
  42, \dodoi{10.3847/1538-4357/aaad0f}

\bibitem[{{Sesana} {et~al.}(2008){Sesana}, {Vecchio}, \&
  {Colacino}}]{Sesana_2008}
{Sesana}, A., {Vecchio}, A., \& {Colacino}, C.~N. 2008, \mnras, 390, 192,
  \dodoi{10.1111/j.1365-2966.2008.13682.x}

\bibitem[{{Sesana} {et~al.}(2009){Sesana}, {Vecchio}, \&
  {Volonteri}}]{Sesana_2009}
{Sesana}, A., {Vecchio}, A., \& {Volonteri}, M. 2009, \mnras, 394, 2255,
  \dodoi{10.1111/j.1365-2966.2009.14499.x}

\bibitem[{{Shankar} {et~al.}(2009){Shankar}, {Weinberg}, \&
  {Miralda-Escud{\'e}}}]{Shankar_2009}
{Shankar}, F., {Weinberg}, D.~H., \& {Miralda-Escud{\'e}}, J. 2009, \apj, 690,
  20, \dodoi{10.1088/0004-637X/690/1/20}

\bibitem[{{Shannon} {et~al.}(2015){Shannon}, {Ravi}, {Lentati}, {Lasky},
  {Hobbs}, {Kerr}, {Manchester}, {Coles}, {Levin}, {Bailes}, {Bhat},
  {Burke-Spolaor}, {Dai}, {Keith}, {Os{\l}owski}, {Reardon}, {van Straten},
  {Toomey}, {Wang}, {Wen}, {Wyithe}, \& {Zhu}}]{Shannon_2015}
{Shannon}, R.~M., {Ravi}, V., {Lentati}, L.~T., {et~al.} 2015, Science, 349,
  1522, \dodoi{10.1126/science.aab1910}

\bibitem[{{Shen} {et~al.}(2008){Shen}, {Greene}, {Strauss}, {Richards}, \&
  {Schneider}}]{Shen_2008}
{Shen}, Y., {Greene}, J.~E., {Strauss}, M.~A., {Richards}, G.~T., \&
  {Schneider}, D.~P. 2008, \apj, 680, 169, \dodoi{10.1086/587475}

\bibitem[{{Soltan}(1982)}]{Soltan_1982}
{Soltan}, A. 1982, \mnras, 200, 115

\bibitem[{{Surace} {et~al.}(1998){Surace}, {Sanders}, {Vacca}, {Veilleux}, \&
  {Mazzarella}}]{Surace_1998}
{Surace}, J.~A., {Sanders}, D.~B., {Vacca}, W.~D., {Veilleux}, S., \&
  {Mazzarella}, J.~M. 1998, \apj, 492, 116, \dodoi{10.1086/305028}

\bibitem[{{Trainor} \& {Steidel}(2013)}]{Trainor_Steidel_2013}
{Trainor}, R., \& {Steidel}, C.~C. 2013, \apjl, 775, L3,
  \dodoi{10.1088/2041-8205/775/1/L3}

\bibitem[{{Tsai} {et~al.}(2015){Tsai}, {Eisenhardt}, {Wu}, {Stern}, {Assef},
  {Blain}, {Bridge}, {Benford}, {Cutri}, {Griffith}, {Jarrett}, {Lonsdale},
  {Masci}, {Moustakas}, {Petty}, {Sayers}, {Stanford}, {Wright}, {Yan},
  {Leisawitz}, {Liu}, {Mainzer}, {McLean}, {Padgett}, {Skrutskie}, {Gelino},
  {Beichman}, \& {Juneau}}]{Tsai_2015}
{Tsai}, C.-W., {Eisenhardt}, P.~R.~M., {Wu}, J., {et~al.} 2015, \apj, 805, 90,
  \dodoi{10.1088/0004-637X/805/2/90}

\bibitem[{{Ueda} {et~al.}(2014){Ueda}, {Akiyama}, {Hasinger}, {Miyaji}, \&
  {Watson}}]{Ueda_2014}
{Ueda}, Y., {Akiyama}, M., {Hasinger}, G., {Miyaji}, T., \& {Watson}, M.~G.
  2014, \apj, 786, 104, \dodoi{10.1088/0004-637X/786/2/104}

\bibitem[{{Veilleux} {et~al.}(2002){Veilleux}, {Kim}, \&
  {Sanders}}]{Veilleux_2002}
{Veilleux}, S., {Kim}, D.-C., \& {Sanders}, D.~B. 2002, \apjs, 143, 315,
  \dodoi{10.1086/343844}

\bibitem[{{Wu} {et~al.}(2015){Wu}, {Wang}, {Fan}, {Yi}, {Zuo}, {Bian}, {Jiang},
  {McGreer}, {Wang}, {Yang}, {Yang}, {Thompson}, \& {Beletsky}}]{Wu_2015}
{Wu}, X.-B., {Wang}, F., {Fan}, X., {et~al.} 2015, \nat, 518, 512,
  \dodoi{10.1038/nature14241}

\bibitem[{{Yu}(2002)}]{Yu_2002}
{Yu}, Q. 2002, \mnras, 331, 935, \dodoi{10.1046/j.1365-8711.2002.05242.x}

\bibitem[{{Yu} \& {Tremaine}(2002)}]{Yu_Tremaine_2002}
{Yu}, Q., \& {Tremaine}, S. 2002, \mnras, 335, 965,
  \dodoi{10.1046/j.1365-8711.2002.05532.x}

\end{thebibliography}

\end{document}